\documentclass[acmsmall,screen,nonacm]{acmart}
\setcopyright{none}
\usepackage{graphicx}
\usepackage{amsmath}
\usepackage{booktabs}
\usepackage{url}
\usepackage{xcolor}
\usepackage{enumitem}
\newcommand{\unmarkedfootnote}[1]{%
  \begingroup
  \renewcommand{\thefootnote}{}%
  \begin{NoHyper}\footnotetext{#1}\end{NoHyper}%
  \addtocounter{footnote}{-1}%
  \endgroup
}
\usepackage{listings}
\usepackage{algorithm}
\usepackage{algpseudocode}
\usepackage{multirow}
\usepackage{cleveref}
\usepackage{balance}
\usepackage{tikz}
\usepackage[most,listings]{tcolorbox}

\newcommand{\su}[1]{}  

\newcommand{\revision}[1]{{\color{black}#1}}

\tcbset{textmarker/.style={%
    enhanced, parbox=false,boxrule=0mm,boxsep=0mm,arc=0mm,
    outer arc=0mm,left=1mm,right=1mm,top=7pt,bottom=7pt,
    toptitle=1mm,bottomtitle=1mm,oversize}}
\newtcolorbox{importantBox}{textmarker,
    borderline west={2pt}{0pt}{red},
    colback=red!10!white}

\newcommand{\takeaway}[2]{
  \begin{tcolorbox}[
    left=3mm,
    right=3mm,
    left skip=4pt,
    colback=gray!10,
    colframe=gray!80,
    boxrule=0.5pt,
    leftrule=4pt,
    top=1mm,
    before skip=0mm,
    after skip=0mm
    ]
    \textbf{Takeaway for #1:}
    #2
  \end{tcolorbox}
}

\hypersetup{hidelinks}

\definecolor{MotivBg}{RGB}{247,248,250}
\definecolor{MotivFrame}{RGB}{180,180,180}
\definecolor{LineNumGray}{RGB}{140,140,140}
\definecolor{bggray}{RGB}{245,245,245}
\definecolor{framegray}{RGB}{180,180,180}

\usepackage{pifont}
\newcommand{\bcircle}[1]{\ding{\numexpr181 + #1}}

\setlist[itemize]{leftmargin=*, topsep=2pt, itemsep=2pt}
\setlist[enumerate]{leftmargin=*, topsep=2pt, itemsep=2pt}

\lstdefinestyle{motivation}{
  language=Python,
  basicstyle=\ttfamily\footnotesize,
  numbers=left,
  numberstyle=\tiny\color{LineNumGray},
  numbersep=6pt,
  frame=none,
  tabsize=2,
  breaklines=true,
  showstringspaces=false,
  commentstyle=\itshape\color{gray!70}
}

\newtcblisting{motivationbox}{
  listing only,
  listing options={style=motivation},
  colback=MotivBg,
  colframe=MotivFrame,
  arc=3pt,
  boxrule=0.6pt,
  left=12pt,
  right=6pt,
  top=4pt,
  bottom=4pt,
}

\title[To Run or Not to Run]{To Run or Not to Run: Analyzing the Cost-Effectiveness of Code Execution in LLM-Based Program Repair}
\author{Zhihao Lin}
\authornote{Both authors contributed equally to this research.}
\affiliation{%
  \institution{Beihang University}
  \city{Beijing}
  \country{China}
}
\email{mathieulin@buaa.edu.cn}

\author{Junhua Zhu}
\authornotemark[1]
\affiliation{%
  \institution{Beihang University}
  \city{Beijing}
  \country{China}
}
\email{22371468@buaa.edu.cn}

\author{Mingyi Zhou}
\affiliation{%
  \institution{Beihang University}
  \city{Beijing}
  \country{China}
}
\email{zhoumingyi@buaa.edu.cn}

\author{Xin Wang}
\affiliation{%
  \institution{Wuhan University}
  \city{Wuhan}
  \country{China}
}
\email{xinwang0920@whu.edu.cn}

\author{Zhensu Sun}
\affiliation{%
  \institution{Singapore Management University}
  \city{Singapore}
  \country{Singapore}
}
\email{zssun@smu.edu.sg}

\author{Renyu Yang}
\affiliation{%
  \institution{Beihang University}
  \city{Beijing}
  \country{China}
}
\email{renyuyang@buaa.edu.cn}

\author{David Lo}
\affiliation{%
  \institution{Singapore Management University}
  \city{Singapore}
  \country{Singapore}
}
\email{davidlo@smu.edu.sg}

\author{Li Li}
\authornote{Corresponding author.}
\affiliation{%
  \institution{Beihang University}
  \city{Beijing}
  \country{China}
}
\email{lilicoding@ieee.org}

\renewcommand{\shortauthors}{Lin et al.}

\begin{document}

\begin{abstract}
LLM-based agents for program repair are increasingly build on a ``generate-run-revise'' paradigm, iteratively executing tests to evaluate and refine patches. This execution-based approach has become standard practice in state-of-the-art systems. However, executions can be time-consuming and expensive, yet their impact on these agents remain underexplored. In this paper, we conduct a two-stage empirical study over execution behavior in LLM-based program repair. To characterize execution behavior at scale, we first analyze 7,745 agent traces from SWE-bench leaderboard submissions. \revision{Second, we evaluate 3,000 end-to-end repair attempts across 200 SWE-bench instances and three agents (Claude Code, Codex, and the open-source OpenCode) under four execution paradigms}, which allows for a fine-grained comparison of performance and cost. Our analysis reveals three key observations: \bcircle{1} Code execution is used across all agents and models analyzed, with an average of 8.8 test runs per task. Execution behavior varies substantially across agents and models, with frequency ranging from 2 to 19 per task, and late-stage executions (66--100\% of conversation) consistently achieve higher success rates than early-stage ones (57.9\% average). \bcircle{2} Execution restrictions have little effect on repair success: On commercial agents with SOTA models the resolve-rate gap between \textsc{Prohibited} and \textsc{Unrestricted} is only 1.25pp (not statistically significant, $p > 0.05$). \revision{The corresponding value for open-source OpenCode with Qwen2.5-Coder-32B is $\approx 0$pp, with equivalence holding under both prompt-level and tool-level enforcement of the restriction.} \textsc{Prohibited} saves 56--62\% tokens and 48--54\% wall-clock on Claude Code, and removes the need to maintain per-repository test environments. \bcircle{3} Execution benefit is concentrated rather than uniform. For commercial agents, 54--66\% of cases complete in a single edit, localization accuracy below \textsc{Prohibited} is over 95\%, and 81--100\% of failed cases pass agent-executed validation but fail the official evaluation. \revision{OpenCode with Qwen2.5-Coder-32B shows another failure mode: it retries more frequently and only 11\% of its failed cases pass self-validation.} These patterns suggest that current agents apply execution indiscriminately, paying their cost on instances where it provides little benefit. Execution, therefore, should be treated as a resource with an explicit cost-benefit tradeoff, not a default capability.
\end{abstract}


\maketitle
\unmarkedfootnote{Accepted at ISSTA 2026.}

\section{Introduction}
\label{sec:intro}
Code agents have been widely applied to software engineering tasks, among which automated program repair (APR) stands out as a critical application~\cite{agentless,sweagent,wang2024opendevin}. 
Modern code agents typically follow an iterative workflow for program repair, where code execution plays a central role.
By executing code, agents can reproduce bugs, localize faults, and validate candidate patches through unit tests.
The execution results, such as test outcomes, runtime errors, and logs, serve as feedback that guides subsequent repair steps, enabling agents to progressively improve solution quality.

Despite its utility, code execution is a resource-intensive operation that imposes significant overhead.
On the one hand, it introduces considerable token costs, as agents must generate execution commands, parse execution outputs, and reason over often verbose feedback.
On the other hand, code execution incurs non-trivial latency, requiring agents to wait for compilation, runtime, and test results before proceeding.
For example, executing a comprehensive test suite can take from several minutes to even hours.
\revision{A third, less visible cost is per-repository environment setup: running a project's test suite requires a working runtime with the right language version and all dependencies installed, which in practice means maintaining a tested Docker image for every target repository and release. This is a recurring engineering tax that scales with deployment breadth.}
As code agents are increasingly deployed in real-world and large-scale settings, these costs become significant.
Given the substantial resources consumed by code execution, a fundamental question arises: how important is code execution to the effectiveness of code agents in program repair?

However, the research community has only a limited understanding of the role of code execution in code agents for program repair.
Prior work on such agents has primarily focused on model architectures~\cite{codellama,deepseekcoder}, prompting strategies~\cite{wei2022chain,treeofthoughts}, search algorithms~\cite{zhang2023planning}, or benchmark performance~\cite{jimenez2024swebench}, often treating execution as a necessary but implicit component of the pipeline.
While these studies acknowledge that execution-based feedback is important, there is a lack of systematic investigation into its quantitative impact.
\revision{A closer work is Agentless~\cite{agentless}, which avoids iterative execution by replacing the agent loop with a fixed localization-repair-validation pipeline. However, Agentless removes both the multi-turn loop and execution access at once, so its results cannot isolate execution's contribution. We take a more granular position: the agent loop is valuable, but iterative execution within it may not be.}

As a result, the extent to which execution contributes to agent performance in program repair, and whether its benefits justify its cost, remains unclear.
\revision{To fill this knowledge gap, we conduct the first empirical study that isolates code execution as a single controlled variable within the LLM agent loop for program repair.
Unlike prior agentless~\cite{agentless} work, we hold the agent scaffold fixed (Claude Code~\cite{claudecode2025} with Claude-Sonnet-4.5, Codex~\cite{openaicodex2025} with GPT-5.2-xhigh, and the open-source OpenCode~\cite{opencode2025} with Qwen2.5-Coder-32B~\cite{qwen25coder2025}) and vary only execution access across four paradigms, yielding a controlled measurement of execution's marginal value.}
Specifically, we target execution behaviors that run code artifacts and produce runtime feedback, including test framework invocations (e.g., pytest and python -m unittest) and Python script execution (e.g., python xxx.py).
Through controlled experiments and detailed analysis, we first investigate how execution is conducted by current code agents and then examine how much and why execution influences their effectiveness and efficiency.
In the following sections, we introduce the design of our study and present the findings for each research question.

\smallskip
\noindent \textbf{RQ1: How do code agents conduct code execution?}
Before investigating the effectiveness of execution, we first quantify how agents currently use execution capabilities to provide foundational context for subsequent analyses.
Specifically, we analyze 7,745 publicly available agent traces from the SWE-bench leaderboard, covering four prominent execution-based agents (SWE-agent, OpenHands, LiveSWEAgent, and Mini-SWE-agent), twelve different LLMs (GPT-4, GPT-4o, GPT-5, GPT-5.2, Claude-3-Opus, Claude-3.5-Sonnet, Claude-4-Sonnet, Claude-Opus-4.5, Kimi-K2, Qwen3-480B, Gemini-3-Pro, and DeepSeek-V3.2), and two benchmark datasets (SWE-bench Lite and Verified).
We examine the frequency, timing distribution, and outcomes of test executions conducted by these agents.

\noindent \textit{Findings.}
We observe substantial variation in execution behavior across agents and models.
we analyze 7,745 public traces and observe that execution is used across all agent-model combinations (avg. 8.8 runs per task), with frequency ranges from 2 to 19 per task, and recent models tend to use more executions than older ones. For example, OpenHands with Claude-4-Sonnet averages 18.7 executions per task, while Mini-SWE-agent with GPT-5.2 averages only 2.0.
Late-stage executions (66--100\% of conversation) consistently achieve higher success rates than early-stage ones across all configurations. For instance, OpenHands with Claude-3.5-Sonnet improves from 42\% to 72\%.
The average success rate is 57.9\%, suggesting that agents refine their understanding over time and execute more targeted tests as the repair progresses.

\smallskip
\noindent \textbf{RQ2: What is the impact of code execution on code agents' performance?}
In this RQ, we perform controlled experiments to measure the impact of code execution.
Specifically, we design four experimental settings with progressively increasing access to code execution, ranging from completely restricted to unrestricted access.
We analyze the performance of three code agents: Claude Code (Claude Sonnet 4.5), Codex CLI (GPT-5.2-xhigh), \revision{and the open-source OpenCode (Qwen2.5-Coder-32B-Instruct).}
Due to budget constraints, we conduct experiments on two representative subsets of SWE-bench: the first 100 instances of SWE-bench Lite and the first 100 instances of SWE-bench Verified.

\noindent \textit{Findings.}
The resolve rate gap between restricted and unrestricted execution is small.
For example, Claude Code achieves a 63\% resolve rate without execution access, only 1 percentage point lower than the 64\% achieved with unrestricted execution, while saving 56\% of tokens and 48\% of wall-clock time.
\revision{Across all agent-dataset combinations the difference is not statistically significant ($p > 0.05$, McNemar's test). To rule out data leakage as an explanation, we replicate this on an open-source agent with Qwen2.5-Coder-32B, whose training cutoff predates SWE-bench Verified: OpenCode reaches 10\% resolve rate under both \textsc{Prohibited} and \textsc{Unrestricted} while consuming 3$\times$ fewer tokens without execution. For the bugs studied, agents pay execution's cost even on instances where it confers no benefit.}

\smallskip
\noindent \textbf{RQ3: Under what conditions does code execution benefit code agents?}
Given the findings from RQ2, we investigate \emph{why} execution does not consistently improve outcomes in order to understand when it is beneficial and when it is not.
Specifically, we analyze instances with stable outcomes across modes (Pass$\rightarrow$Pass and Fail$\rightarrow$Fail) to characterize the conditions under which execution adds value.

\noindent \textit{Findings.}
We identify two factors behind the limited benefit on the studied bugs.
First, reproduction execution provides little localization benefit: although 55\% of Claude Code's successful cases use it, localization accuracy stays above 95\% in both modes, and only 48.8\% of reproduction executions produce actionable feedback.
Second, execution feedback often does not correct errors: 54--66\% of commercial-agent cases complete in a single edit regardless of execution access, and 81--100\% of failed cases pass agent-conducted validation but fail the official SWE-bench evaluation. This is a gap between agent-chosen tests and ground-truth validation.
Execution is not inherently unhelpful; its benefit is concentrated on specific instances. The open question is how an agent should decide \emph{when} to invest in execution.

In summary, this paper makes the following contributions:
\begin{itemize}[leftmargin=*, nosep]
    \item The first empirical study that isolates code execution as a single variable within the agent loop for program repair: scaffold held fixed (Claude Code, Codex, and open-source OpenCode), only execution access varied across four paradigms. Where prior agentless work changes scaffold and execution at once, our design measures execution's marginal value alone. Covers 7,745 public traces plus 3,000 controlled repair attempts on 200 SWE-bench instances.
    \item Execution is not consistently beneficial: the \textsc{Prohibited}--\textsc{Unrestricted} resolve-rate gap is 1.25pp on commercial agents \revision{and $\approx 0$pp on the open-source one (none significant, $p>0.05$). We also expose a boundary regime: an open-source model with a 65K-token context (Qwen2.5-Coder-32B) does best with a single well-chosen execution (\textsc{Quota-1}) rather than \textsc{Unrestricted}.} Execution should be treated as a resource with an explicit cost-benefit tradeoff, not a default capability.
    \item We release a reproducible framework\footnote{\url{https://github.com/mathieu0905/To_Run_Or_Not_To_Run}} for limiting code execution of program-repair agents.
\end{itemize}

\section{Background}
\label{sec:background}
In this section, we describe the workflow of code agents and explain how they conduct code execution.

\subsection{Code Agents}
Code agents are LLM-based systems that autonomously perform software engineering tasks by combining reasoning with tool use~\cite{sweagent, codeact, metagpt}.
Unlike simple code completion, these agents wrap an LLM in a control loop that persists state across multiple turns, enabling them to tackle complex, multi-step tasks such as bug fixing, feature implementation, and code refactoring.

A typical code agent workflow combines three core capabilities: \emph{file exploration} (traversing directories and searching for relevant code), \emph{code editing} (applying patches via diffs or block replacements), and \emph{code execution} (running shell commands to invoke build systems, linters, and test suites).
Representative agents include SWE-agent~\cite{sweagent}, OpenHands~\cite{wang2024opendevin}, and commercial CLIs such as Claude Code~\cite{claudecode2025} and Codex~\cite{openaicodex2025}.

For program repair tasks, agents typically follow an iterative loop~\cite{sweagent, chatrepair}:
\begin{center}
\texttt{inspect code} $\rightarrow$ \texttt{propose patch} $\rightarrow$ \texttt{run tests} $\rightarrow$ \texttt{revise}
\end{center}
This loop makes test execution the primary feedback channel: agents run tests to validate their patches and use test outputs to guide subsequent revisions.
A notable alternative is Agentless~\cite{agentless}, which uses a two-phase pipeline (localization followed by repair) without iterative execution, yet achieves competitive results on SWE-bench.
This raises a key question: how much value does the execution-heavy agentic paradigm actually provide?

\subsection{Code Execution by Code Agents}
Code execution is the process by which agents run code artifacts and observe runtime feedback.
The typical execution cycle proceeds as follows: the agent first \emph{writes a command} (e.g., \texttt{pytest tests/test\_foo.py}), the command is then \emph{executed} in a shell environment with access to the project's dependencies, the agent \emph{observes the results} (test outputs, error messages, and stack traces), and based on these results, decides whether to \emph{revise the patch} or submit.

This execution loop incurs substantial costs across multiple dimensions.
\emph{Wall-clock time}: agents must wait for execution to complete, with test suites taking seconds to minutes and creating latency in the repair loop.
\emph{Token consumption}: execution outputs, including verbose test logs and stack traces, are fed back to the LLM, consuming context window capacity and increasing API costs.
\emph{Environment overhead}: each execution requires a properly configured environment with dependencies, databases, and external services.
These costs compound when agents engage in trial-and-error behavior, repeatedly executing tests without making substantive progress.
Recent work has begun addressing efficiency: Peng et al.~\cite{peng2025more} show that limiting interaction turns can reduce costs by 24--68\% with minimal impact on solve rates.
However, turn-level budgets treat all interactions equally, obscuring the distinction between cheap operations (reading files) and expensive ones (running tests).
Our study focuses specifically on execution costs, providing a fine-grained analysis of when execution helps and when it merely adds overhead.

\section{Experimental Setup}
\label{sec:method}
In this section, we introduce the experimental setup of this study, including the settings for code execution, benchmark, agents, evaluation metrics, and implementation details.
Our study is guided by three research questions:
\begin{itemize}[leftmargin=*]
    \item \textbf{RQ1}: How do code agents conduct code execution?
    \item \textbf{RQ2}: What is the impact of code execution on code agents' performance?
    \item \textbf{RQ3}: \revision{Under what conditions does code execution benefit code agents, and why is its impact not consistently positive?}
\end{itemize}

\subsection{Experimental Settings for Code Execution}
\label{sec:exec-paradigms}
\su{please update the content based on the new title -- DONE}
\su{move this as the first subsection -- DONE: now first subsection}
\su{introduce the execution of what code are targeted by this study at the beginning (refer to the comments of the Terminology subsection) -- DONE: see paragraph below}
\su{then describe the four settings which limits the execution from different levels -- DONE}
\su{The name of each setting can be more straightforward: Prohibited, Quota-limited, Budget-guided, Unrestricted -- DONE: renamed all}

\su{explain this at the beginning of `Experimental Settings for Code Execution' -- DONE: merged Terminology content here}\su{mainly introduce the targeted execution types and their commands. -- DONE}
In this study, we focus on \emph{code execution}, which refers to operations that run code artifacts and produce runtime feedback. This includes test framework invocations (\texttt{pytest}, \texttt{python -m unittest}, \texttt{python manage.py test}, \texttt{tox}, \texttt{nosetests}) and Python script execution (\texttt{python xxx.py}).
Other commands such as \texttt{ls}, \texttt{cat}, \texttt{grep}, and \texttt{find} are exploratory in nature and are not restricted.

We study four execution paradigms that form a spectrum from no execution to unlimited access.
\textsc{Prohibited} mode restricts access to project-specific runtime environments: the agent is instructed via prompt to avoid running test frameworks or project-specific scripts, and project dependencies (e.g., Django, Flask, Sympy) are not installed.
Since this is a soft constraint via prompting, agents occasionally still attempt to execute tests or scripts; however, these attempts fail due to missing dependencies and do not provide useful runtime feedback.
A basic Python interpreter remains available for simple commands such as \texttt{python -c "print(1+1)"}.
This design reflects a realistic ``read-only analysis'' scenario where developers reason about code without setting up the full project environment.
\begin{tcolorbox}[left=2mm, right=2mm, top=1mm, bottom=1mm, colback=gray!5, colframe=gray!60, boxrule=0.4pt]
\small\textbf{Prompt:} \texttt{Do not run pytest, unittest, or project test scripts. Project dependencies are not installed.}
\end{tcolorbox}

\textsc{Quota-Limited} mode allows execution with a point budget, where the agent self-estimates costs before each run and test-framework invocations are treated as expensive operations.
We evaluate two budget levels: K=1 (minimal execution) and K=3 (moderate execution), yielding five experimental conditions from four paradigms.
\begin{tcolorbox}[left=2mm, right=2mm, top=1mm, bottom=1mm, colback=gray!5, colframe=gray!60, boxrule=0.4pt]
\small\textbf{Prompt:} \texttt{You have a budget of K test run(s). Unused budget is wasted opportunity!}\\
\textbf{Cost Table:} \texttt{pytest}/\texttt{unittest}/\texttt{python manage.py test} = 1.0 point; \texttt{python script.py} = 0.3 point.
\end{tcolorbox}

\textsc{Budget-Guided} mode permits unrestricted execution but prompts the agent to consider whether each run is worth its cost, testing whether cost awareness alone can reduce unnecessary execution.
\begin{tcolorbox}[left=2mm, right=2mm, top=1mm, bottom=1mm, colback=gray!5, colframe=gray!60, boxrule=0.4pt]
\small\textbf{Prompt:} \texttt{You CAN run tests and scripts, but each execution has a cost. Goal: Fix the bug correctly while being mindful of execution costs.}
\end{tcolorbox}

At the other extreme, \textsc{Unrestricted} mode allows unlimited execution, enabling the typical ``generate--run--revise'' loop that characterizes most current agents. No execution constraints are specified in the prompt.

This configuration yields \revision{600 unique agent-instance combinations (200 instances $\times$ 3 agents), totaling 3,000 end-to-end repair attempts} across five execution modes.

\subsection{Benchmark}
We study execution behavior in the context of automated program repair on SWE-bench~\cite{swebench}, a benchmark of real-world GitHub repositories and their issues.
Each instance provides a repository snapshot, a problem statement, and a test-based evaluation protocol.
Due to budget constraints, we use two commonly used variants: SWE-bench Lite (300 instances) and SWE-bench Verified (500 instances), selecting the first 100 instances from each dataset in their canonical ordering.
This yields 200 instances spanning diverse repositories including Django, Flask, Requests, and Sympy.
The task requires an agent to understand the bug from the problem statement, locate relevant code in the repository, generate a patch, and optionally validate the fix via test execution.
The official SWE-bench harness evaluates predictions by applying the patch in a clean container and running instance-specific tests.

\subsection{Models and Agents}
\su{also introduce the agents and models in RQ1 here -- DONE: added two-stage design description}
\su{we can say that we first do the observation on xx and then due to limited budget we experiment on xx -- DONE}
Our study adopts a two-stage design.
In the first stage (RQ1), we analyze execution behavior across a broad range of publicly available agent traces from the SWE-bench leaderboard, covering four prominent agents (SWE-agent, OpenHands, LiveSWEAgent, and Mini-SWE-agent) and twelve LLMs (GPT-4, GPT-4o, GPT-5, GPT-5.2, Claude-3-Opus, Claude-3.5-Sonnet, Claude-4-Sonnet, Claude-Opus-4.5, Kimi-K2, Qwen3-480B, Gemini-3-Pro, and DeepSeek-V3.2).
In the second stage (RQ2--RQ3), \revision{we conduct controlled experiments with three agents: Claude Code~\cite{claudecode2025} (Claude Sonnet 4.5), Codex~\cite{openaicodex2025} (GPT-5.2-xhigh), and the open-source OpenCode~\cite{opencode2025} with Qwen2.5-Coder-32B-Instruct~\cite{qwen25coder2025} served via vLLM~\cite{vllm2024}.\footnote{Precise CLI and model versions for reproducibility: Claude Code v1.0.16 (Anthropic \texttt{claude-sonnet-4-5}); Codex v0.1.2025062301 (OpenAI \texttt{gpt-5.2} with reasoning effort \texttt{xhigh}); OpenCode v1.4.0 with tensor-parallel vLLM serving \texttt{Qwen/Qwen2.5-Coder-32B-Instruct}. For brevity, unless otherwise specified, we refer to these fixed agent--model configurations as \emph{Claude Code}, \emph{Codex}, and \emph{OpenCode}, respectively, in the following.}
All agents operate in the SWE-bench containerized environment.
Claude Code and Codex represent state-of-the-art commercial offerings; OpenCode with Qwen2.5-Coder-32B provides an open-source, open-weight model baseline that mitigates data contamination concerns (see \Cref{sec:discussion}).}

\subsection{Metrics}\su{I think the official name from SWE-Bench is Resolve Rate, not Pass Rate. -- DONE: changed to resolve rate}
We record interaction traces and compute several metrics to characterize agent behavior and outcomes.
For execution behavior, we measure \emph{execution frequency} as the average number of test executions per task, \emph{timing distribution} as the percentage of executions occurring in each conversation stage (Early: 0--33\%, Middle: 33--66\%, Late: 66--100\%), and \emph{execution outcome} as the success or failure rate of test executions.
For effectiveness and cost, the primary outcome is \emph{resolve rate}, which indicates the proportion of instances where the generated patch passes the official SWE-bench evaluation.\su{proportion -- DONE}
We measure \emph{token consumption} (input and output tokens) to reflect API cost, and \emph{wall-clock time} to capture end-to-end runtime.
To ensure fair time comparisons, all execution modes for the same instance run concurrently on identical hardware.
For understanding execution impact, we measure \emph{localization accuracy} using Hit (at least one edited file matches ground truth) and Recall (proportion of ground truth files edited), \emph{single-edit ratio} as the percentage of instances with only one code edit and no subsequent modifications, \emph{post-execution modification ratio} as the percentage of instances where code is modified after test execution, and \emph{actionable feedback ratio} as the percentage of reproduction executions that provide useful localization information.

\subsection{Implementation Details}\su{Report these: how much is costed for this study, how do we access the two agents -- DONE: added CLI access info}
\label{sec:implementation}

\paragraph{Experimental Scale.}
Our evaluation comprises \revision{\textbf{3,000 end-to-end agent runs}: 200 SWE-bench instances (100 from Lite, 100 from Verified) $\times$ 3 agents (Claude Code, Codex, OpenCode)} $\times$ 5 execution modes (Prohibited, Quota-Limited $K{=}1$, Quota-Limited $K{=}3$, Budget-Guided, Unrestricted).
Each run involves a complete repair attempt: checking out the repository, reading the issue, generating and applying patches, and running the official SWE-bench evaluation harness.
Critically, all execution modes are evaluated on \emph{exactly the same} 200 instances; this paired design controls for instance-level difficulty variation, enabling reliable statistical analysis through paired comparisons.

\revision{We access all agents via their command-line interfaces: Claude Code via \texttt{claude -p}, Codex via \texttt{codex exec}, and OpenCode via \texttt{opencode}.}
All modes share the same prompt structure, consisting of task description, repository information, problem statement, execution constraint, and output format.
Only the execution policy differs across modes, isolating execution access as the primary experimental variable.
The point budget serves as a guideline rather than a hard limit; we report outcomes by assigned budget regardless of compliance, allowing us to study how agents respond to cost awareness even when not strictly enforced.
The final patch is extracted via \texttt{git diff} and evaluated using the official SWE-bench harness, which applies the patch in a clean container and runs the instance-specific evaluation script.

\section{Evaluation}
\label{sec:eval}

In this section, we present the results of our empirical study, organized around three research questions that examine how agents use execution, the impact on repair effectiveness and cost, and why execution feedback has limited benefits.

\subsection{RQ1: How Do Code Agents Conduct Code Execution?}
\label{sec:rq1}
Before examining the effectiveness of execution, we first quantify how frequently execution is conducted in state-of-the-art coding agents. We analyzed 7,745 publicly available agent traces (i.e., the complete logs of agent-environment interactions during repair attempts) from the SWE-bench leaderboard, covering four prominent execution-based agents (SWE-agent, OpenHands, LiveSWEAgent, and Mini-SWE-agent), twelve different LLMs (GPT-4, GPT-4o, GPT-5, GPT-5.2, Claude-3-Opus, Claude-3.5-Sonnet, Claude-4-Sonnet, Claude-Opus-4.5, Kimi-K2, Qwen3-480B, Gemini-3-Pro, and DeepSeek-V3.2), and two benchmark datasets (SWE-bench Lite and Verified).

For each code execution, we record two dimensions.
First, we record its \emph{timing}, defined as the normalized position in the agent conversation where 0.0 represents the start and 1.0 represents the end. We categorize timing into three stages: Early (0--33\%), Middle (33--66\%), and Late (66--100\%).
Second, we record its \emph{outcome}, classified as either Success or Failure. A successful execution is one where pytest reports ``passed'' without any ``FAILED'' messages, unittest outputs ``OK'', or the command returns exit code 0. A failed execution includes test assertion failures, Python exceptions, or non-zero exit codes.
Table~\ref{tab:cross-agent-exec} presents the test execution analysis across all agent-model combinations. Our observations are as follows:

\begin{table*}[t]
  \centering
  \small
  \setlength{\tabcolsep}{3pt}
  \caption{Test execution analysis on 7,745 SWE-bench public traces. \#Traces is the number of leaderboard submissions for that (agent, model) pair. \emph{Exec/Task} is the mean number of test-framework invocations (\texttt{pytest}, \texttt{python -m unittest}, \texttt{python manage.py test}, \texttt{tox}, \texttt{nosetests}) per instance, averaged over all traces for that pair. Early/Middle/Late = percentage of executions occurring in each conversation stage (0--33\%, 33--66\%, 66--100\% of turns before the final patch). Success/Failure = percentage of test executions whose exit status indicates all tests passed vs.\ at least one test failed or errored.}
  \label{tab:cross-agent-exec}
  \begin{tabular}{llccccccc}
    \toprule
    & & \multicolumn{2}{c}{Stats} & \multicolumn{3}{c}{Stage} & \multicolumn{2}{c}{Outcome} \\
    \cmidrule(lr){3-4} \cmidrule(lr){5-7} \cmidrule(lr){8-9}
    Agent & Model & \#Traces & Exec/Task & \%Early & \%Middle & \%Late & \%Success & \%Failure \\
    \midrule
    \multicolumn{9}{l}{\textit{SWE-bench Lite}} \\
    SWE-agent & GPT-4 & 254 & 3.2 & 42.4\% & 28.1\% & 29.6\% & 46.1\% & 53.9\% \\
    SWE-agent & Claude-3.5-Sonnet & 259 & 6.3 & 23.5\% & 33.0\% & 43.4\% & 38.7\% & 61.3\% \\
    SWE-agent & GPT-4o & 259 & 7.4 & 32.3\% & 33.4\% & 34.2\% & 36.7\% & 63.3\% \\
    OpenHands & Claude-3.5-Sonnet & 291 & 6.0 & 19.8\% & 33.7\% & 46.4\% & 57.8\% & 42.2\% \\
    \midrule
    \multicolumn{9}{l}{\textit{SWE-bench Verified}} \\
    SWE-agent & GPT-4 & 400 & 3.3 & 42.7\% & 27.9\% & 29.4\% & 45.8\% & 54.2\% \\
    SWE-agent & Claude-3-Opus & 366 & 3.7 & 35.8\% & 31.2\% & 32.9\% & 45.3\% & 54.7\% \\
    SWE-agent & Claude-3.5-Sonnet & 434 & 6.9 & 24.8\% & 34.5\% & 40.8\% & 40.3\% & 59.7\% \\
    SWE-agent & GPT-4o & 425 & 7.5 & 31.2\% & 33.2\% & 35.6\% & 30.4\% & 69.6\% \\
    OpenHands & Claude-3.5-Sonnet & 488 & 6.7 & 22.7\% & 33.4\% & 43.9\% & 58.6\% & 41.4\% \\
    OpenHands & Claude-4-Sonnet & 500 & 18.7 & 15.4\% & 30.9\% & 53.7\% & 70.4\% & 29.6\% \\
    OpenHands & Kimi-K2 & 500 & 15.5 & 11.7\% & 34.9\% & 53.3\% & 66.4\% & 33.6\% \\
    OpenHands & Qwen3-480B & 500 & 18.7 & 12.4\% & 31.7\% & 55.9\% & 70.3\% & 29.7\% \\
    OpenHands & GPT-5 & 479 & 4.1 & 11.5\% & 38.6\% & 49.9\% & 68.1\% & 31.9\% \\
    LiveSWEAgent & Gemini-3-Pro & 500 & 12.6 & 28.7\% & 38.2\% & 33.2\% & 77.8\% & 22.2\% \\
    LiveSWEAgent & Claude-Opus-4.5 & 500 & 15.9 & 23.6\% & 40.1\% & 36.3\% & 79.3\% & 20.7\% \\
    Mini-SWE-agent & Claude-Opus-4.5 & 495 & 10.5 & 12.6\% & 38.3\% & 49.1\% & 78.7\% & 21.3\% \\
    Mini-SWE-agent & Gemini-3-Pro & 495 & 8.6 & 23.1\% & 36.6\% & 40.3\% & 71.8\% & 28.2\% \\
    Mini-SWE-agent & DeepSeek-V3.2 & 485 & 10.5 & 14.2\% & 34.9\% & 50.9\% & 59.9\% & 40.1\% \\
    Mini-SWE-agent & GPT-5.2 & 115 & 2.0 & 1.8\% & 21.9\% & 76.3\% & 58.0\% & 42.0\% \\
    \bottomrule
  \end{tabular}
\end{table*}

\subsubsection{Statistical Analysis}
First, code execution is widely adopted across all agent-model combinations.
On average, agents perform 8.8 test executions per task.
For models such as Claude-Opus-4.5 in LiveSWEAgent, the execution count can reach 15.9 per instance.
Second, execution behavior varies substantially across agents and models.
Recent models, with the exception of GPT-5.2, tend to use more executions than older models.
For example, OpenHands with Claude-4-Sonnet averages 18.7 executions per task, while Mini-SWE-agent with GPT-5.2 averages only 2.0, representing a 9$\times$ difference.

\subsubsection{Timing Analysis}
Late-stage execution (66--100\% of conversation) is the most common pattern across configurations.
For example, OpenHands with Qwen3-480B concentrates 55.9\% of executions in the late stage.
However, older models like GPT-4 show a different pattern, with more executions in the early stage.
SWE-agent with GPT-4 performs 42.4\% of executions in the early stage, compared to only 29.6\% in the late stage.
Notably, some state-of-the-art models show minimal early-stage execution.
For example, GPT-5.2 in Mini-SWE-agent has only 1.8\% early-stage executions, and Kimi-K2 in OpenHands has 11.7\%.
This pattern may suggest that these models understand the issue well from the problem description alone, reducing the need for reproduction before editing.

\subsubsection{Outcome Analysis}
The overall execution success rate is moderately high, with an average of 57.9\% across all configurations.
Success rates range from 30.4\% for SWE-agent with GPT-4o to 79.3\% for LiveSWEAgent with Claude-Opus-4.5.
Some agent-model combinations achieve particularly high success rates.
For example, LiveSWEAgent with Claude-Opus-4.5 and Mini-SWE-agent with Claude-Opus-4.5 both exceed 78\%.
Among the failures, TestFailure (assertion failures), TestError (collection or setup errors), and Python exceptions (such as ModuleNotFoundError, AttributeError, and TypeError) are the most common types.
Due to space constraints, the complete failure categorization is available in our artifacts.

\begin{figure}[h]
    \centering
    \includegraphics[width=\columnwidth]{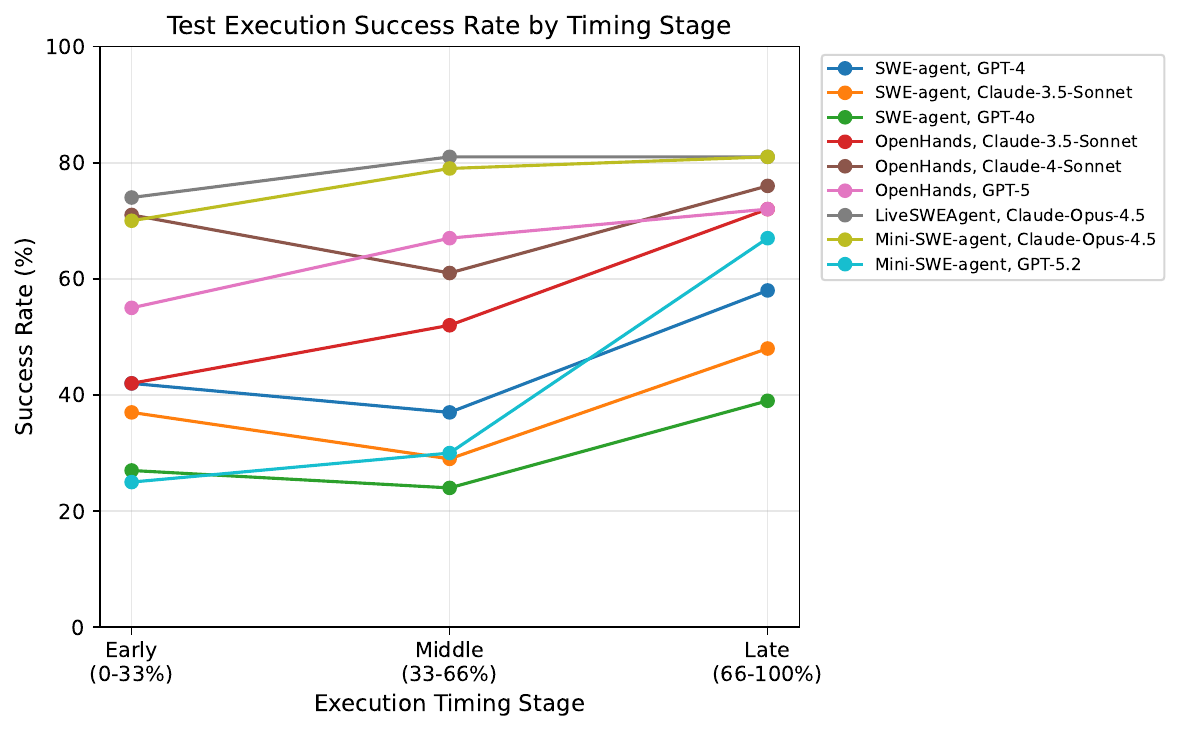}
    \caption{Test execution success rate by timing stage. Late-stage executions (66--100\% of conversation) consistently achieve higher success rates than early-stage ones across all agent-model combinations.}
    \Description{A line chart with three data points (Early, Middle, Late) on the x-axis and success rate percentage on the y-axis. Multiple lines represent different agent-model combinations. All lines show an upward trend from Early to Late stages, with success rates improving from approximately 25-50\% in Early stage to 55-80\% in Late stage.}
    \label{fig:timing-outcome}
\end{figure}

As shown in Figure~\ref{fig:timing-outcome}, tests executed in late stages (66--100\% of conversation) consistently achieve higher success rates than early-stage tests across all configurations.
For example, OpenHands with Claude-3.5-Sonnet improves from 42\% (early) to 72\% (late), and Mini-SWE-agent with GPT-5.2 improves from 25\% to 67\%.
This pattern suggests that agents refine their understanding over time and execute more targeted, better-formed tests as the repair progresses.

\vspace{1em}
\takeaway{RQ1}{
\begin{itemize}[leftmargin=*, nosep]
  \item Code execution is commonly used across all analyzed agent-model combinations, with an average of 8.8 test runs per task.
  \item Across 7,745 traces spanning 4 agents and 12 models, execution behavior varies substantially: frequency ranges from 2 to 19 per task, and timing patterns differ by agent architecture.
  \item The average success rate is 57.9\%, with late-stage executions consistently achieving higher success rates than early-stage ones (e.g., 42\% $\rightarrow$ 72\% for OpenHands with Claude-3.5-Sonnet).
\end{itemize}
}

\subsection{RQ2: Effectiveness and Cost Analysis}
\label{sec:rq2}
In this RQ, we systematically evaluate the impact of execution access on repair effectiveness and cost.
Specifically, we evaluate three agents (Claude Code, Codex, and the open-source OpenCode with Qwen2.5-Coder-32B) on two benchmarks (SWE-bench Lite and Verified) under four execution paradigms (\textsc{Prohibited}, \textsc{Quota-Limited}, \textsc{Budget-Guided}, and \textsc{Unrestricted}), realised as five configurations (\textsc{Quota-Limited} is instantiated at $K{=}1$ and $K{=}3$).
We measure resolve rate, defined as the percentage of instances where the generated patch passes all tests in the official SWE-bench evaluation harness, and record token consumption and wall-clock time following the methodology outlined in Section~\ref{sec:method}.
Additionally, we assess statistical significance between resolve rates using McNemar's test~\cite{mcnemar1947} and equivalence testing (Two One-Sided Tests, TOST~\cite{schuirmann1987}, with equivalence margin $\delta = 5$ percentage points, or pp).
The results are presented in Table~\ref{tab:pass-rate-full}, where we mark with $\dagger$ cases where \textsc{Prohibited} performance is within 3pp of \textsc{Unrestricted}.
\revision{The \textsc{Prohibited}--\textsc{Unrestricted} gap stays within $\pm 5$pp across the six headline cells and averages 1.25pp on commercial agents. We present the effectiveness result and its data-leakage check first (\Cref{tab:pass-rate-full,tab:ci-full}), then verify that it does not depend on whether agents fully obey the prompt-level constraint (\Cref{tab:compliance,tab:compliant-subset}), and end with cost (\Cref{tab:cost-full,tab:cost-comparison}).}

\subsubsection{Effectiveness Analysis}

Table~\ref{tab:pass-rate-full} presents the complete resolve rate comparison across all agents, benchmarks, and execution modes. The central observation is that the gap between \textsc{Prohibited} and \textsc{Unrestricted} is remarkably small: averaged across the four commercial-agent cells the difference is only 1.25pp, and averaged across all six cells (including OpenCode) it is $-0.83$pp.

\begin{table*}[!ht]
  \centering
  \footnotesize
  \setlength{\tabcolsep}{4pt}
  \caption{Resolve rate (\%) across all execution modes. Bold indicates best performance per row. $\dagger$: \textsc{Prohibited} performance is within 3pp of \textsc{Unrestricted}. None of the differences are statistically significant ($p > 0.05$, McNemar's test).}
  \label{tab:pass-rate-full}
  \begin{tabular}{llccccc}
    \toprule
    Agent & Benchmark & Prohibited & Quota-1 & Quota-3 & Budget-Guided & Unrestricted \\
    \midrule
    \multirow{2}{*}{Claude Code}
      & Lite & 63.0$^\dagger$ & 61.0 & 62.0 & 63.0 & \textbf{64.0} \\
      & Verified & 64.0$^\dagger$ & 64.0 & 65.0 & \textbf{67.0} & \textbf{67.0} \\
    \midrule
    \multirow{2}{*}{Codex}
      & Lite & \textbf{74.0}$^\dagger$ & 68.0 & 69.0 & 71.0 & 73.0 \\
      & Verified & 73.0$^\dagger$ & 72.0 & 73.0 & 71.0 & \textbf{75.0} \\
    \midrule
    \multirow{2}{*}{\revision{OpenCode}}
      & \revision{Lite} & \revision{7.0} & \revision{\textbf{14.0}} & \revision{7.0} & \revision{9.0} & \revision{6.0} \\
      & \revision{Verified} & \revision{13.0} & \revision{\textbf{17.0}} & \revision{11.0} & \revision{13.0} & \revision{14.0} \\
    \bottomrule
  \end{tabular}
\end{table*}

Several patterns emerge from these results.
First, we do not observe a monotonic relationship between execution access and repair success. For Codex on Lite, \textsc{Prohibited} achieves the highest resolve rate (74.0\%) among all modes, outperforming \textsc{Unrestricted} (73\%) by 1.0 percentage point.
Second, the \textsc{Quota-Limited} modes often perform \emph{worse} than \textsc{Prohibited}.
For example, Codex on Lite achieves 74.0\% under \textsc{Prohibited} but drops to 68.0\% under \textsc{Quota-1} and 69.0\% under \textsc{Quota-3}, a decrease of up to 6 percentage points.
This suggests that partial execution access may be counterproductive, possibly because limited feedback is insufficient for effective iteration; we revisit this observation in RQ3.
Third, the open-source OpenCode agent (Qwen2.5-Coder-32B) exhibits the same equivalence pattern on a non-commercial stack: despite much lower absolute resolve rates---expected given the 32B parameter count versus frontier models---the \textsc{Prohibited} vs.\ \textsc{Unrestricted} gap is within $\pm 1$pp on both benchmarks (Lite: 7.0\% vs.\ 6.0\%; Verified: 13.0\% vs.\ 14.0\%), averaging 0pp.
Since Qwen2.5-Coder-32B-Instruct's training cutoff predates \textsc{Verified}, this extends the equivalence to a regime with limited exposure to SWE-bench solutions (see \Cref{sec:threats} for a dedicated training-data discussion).
Overall, these results indicate that the impact of code execution on repair effectiveness is limited.

\revision{OpenCode does, however, exhibit one boundary behaviour worth flagging: its highest-resolve mode is \textsc{Quota-1} (Lite 14.0\%, Verified 17.0\%; paired-bootstrap 90\% CI $[+5.9, +15.3]$pp vs.\ \textsc{Prohibited} on Lite), driven by the non-empty-patch rate collapsing from 74/100 (Lite) and 76/100 (Verified) under \textsc{Quota-1} to 50/100 and 56/100 under \textsc{Unrestricted} as test output crowds the 65K context. We treat this as a small-model boundary effect.}

To rigorously assess these differences, Table~\ref{tab:ci-full} reports both 95\% Wilson confidence intervals (preferred for proportions with small sample sizes) and paired McNemar tests for \textsc{Prohibited} vs.\ \textsc{Unrestricted}.
All six agent-benchmark combinations show substantially overlapping Wilson intervals and non-significant McNemar results ($p > 0.05$), confirming that observed differences are within statistical noise.
The discordant pairs ($b$ = \textsc{Prohibited} succeeds / \textsc{Unrestricted} fails; $c$ = reverse) are few in number and near-symmetric across cells (5:6, 6:9, 4:3, 1:3), consistent with agent stochasticity rather than a systematic execution benefit.
Formal TOST equivalence with $\delta = 5$pp holds for Codex/Verified (90\% CI: [$-$0.8pp, $+$4.8pp]); the remaining five cells keep point estimates within the $\pm 5$pp band but do not meet formal equivalence at $n=100$, a power-vs.-sample-size limitation we revisit in \Cref{sec:threats}.
In summary, the statistical results confirm that, for the bugs studied, execution access does not produce a reliably positive effect on repair outcomes.

\begin{table}[h]
  \centering
  \footnotesize
  \setlength{\tabcolsep}{3pt}
  \caption{Resolve rates with 95\% Wilson CIs and paired McNemar test (\textsc{Prohibited} vs.\ \textsc{Unrestricted}). $b$ = Prohib.\ succeeds / Unrestr.\ fails; $c$ = reverse; all $p > 0.05$.}
  \label{tab:ci-full}
  \label{tab:mcnemar}
  \begin{tabular}{llccccc}
    \toprule
    Agent & Bench. & Prohib.\ (CI) & Unrestr.\ (CI) & $b$ & $c$ & $p$ \\
    \midrule
    Claude Code & Lite & 63.0 [53.2,71.8] & 64.0 [54.2,72.7] & 5 & 6 & 1.000 \\
    Claude Code & Ver. & 64.0 [54.2,72.7] & 67.0 [57.3,75.4] & 6 & 9 & 0.607 \\
    Codex & Lite & 74.0 [64.5,81.8] & 73.0 [63.6,80.7] & 4 & 3 & 1.000 \\
    Codex & Ver. & 73.0 [63.6,80.7] & 75.0 [65.7,82.5] & 1 & 3 & 0.625 \\
    \revision{OpenCode} & \revision{Lite} & \revision{7.0 [3.4,13.7]} & \revision{6.0 [2.8,12.5]} & \revision{4} & \revision{3} & \revision{1.000} \\
    \revision{OpenCode} & \revision{Ver.} & \revision{13.0 [7.8,21.0]} & \revision{14.0 [8.5,22.1]} & \revision{4} & \revision{5} & \revision{1.000} \\
    \bottomrule
  \end{tabular}
\end{table}

As a sanity check on whether \textsc{Prohibited}'s strong performance reflects \emph{verbatim} memorisation rather than reasoning, we compared the patches produced in the two modes: if memorisation were the dominant mechanism, patches should largely coincide. Using \url{difflib.SequenceMatcher} on normalised patches, we find the identical-patch rate is 24\% on Claude Code, 1\% on Codex, and \revision{14\% on OpenCode}, with mostly same-file / different-code pairs (Claude Code 15\%, Codex 27\%, \revision{OpenCode 24\%}); average patch similarity is 42--60\%. This is limited but non-trivial overlap, and it rules out \emph{rote recitation} as the dominant mechanism. \revision{The OpenCode result is particularly informative: Qwen2.5-Coder-32B-Instruct's training cutoff predates \textsc{Verified}, yet the \textsc{Prohibited}--\textsc{Unrestricted} equivalence still holds on that benchmark, further constraining the role of training-data overlap.}

\subsubsection{Hard-Constraint Verification}\label{sec:compliance}
Because \textsc{Prohibited} is enforced at the prompt level, agents may occasionally still attempt to run tests(as shown in \Cref{tab:compliance}). This raises a  concern that whether these unintended executions explain \textsc{Prohibited}'s strong performance. We address this in two steps: (i) measure what each configuration actually delivers, and show that the unintended executions in \textsc{Prohibited} do not carry enough usable feedback to explain its resolve rate, and (ii) re-verify on the hard-constraint.

\emph{(i) What each configuration actually delivers.}
\Cref{tab:compliance} splits each cell into four per-instance averages: \emph{Attempted} test-framework invocations, \emph{Env-Err} attempts blocked by missing modules or dependency errors before reaching the test stage, \emph{Completed} attempts that reach the test stage and produce non-empty output, and \emph{Actionable} attempts that produce a concrete pass/fail signal the agent could use to guide the next edit. The four columns separate intent (\emph{Attempted}) from usable feedback (\emph{Actionable}), which diverge sharply under \textsc{Prohibited}: missing dependencies turn most attempts into dead ends.
Under \textsc{Prohibited}, Codex essentially obeys the prompt outright (0.00--0.01 attempts per instance); OpenCode attempts 0.77 on Lite but drops to zero on Verified; Claude Code is the loudest violator, attempting 0.76--0.79 times per instance. However, environment errors absorb about a quarter of them before any test output is produced, leaving about 0.55--0.59 completed runs per instance; of those, only 0.36--0.39 yield a concrete pass/fail signal the agent could act on. The effective rate of usable unintended feedback is thus well under half the per-instance quota of \textsc{Quota-1} and roughly an eighth of the \textsc{Unrestricted} rate (5.30 on Lite, 4.98 on Verified). The same pattern is even more pronounced on OpenCode (0.14 and 0.00 actionable per instance) and essentially zero on Codex.

\begin{table*}[h]
  \centering
  \caption{Per-mode compliance: each cell reports \emph{Attempted/Env-Err/Completed/Actionable} as per-instance averages over 100 instances. \emph{Attempted} = test-framework invocations launched; \emph{Env-Err} = blocked by missing modules or dependency errors; \emph{Completed} = reached the test stage; \emph{Actionable} = produced a pass/fail signal (subset of Completed).}
  \label{tab:compliance}
  \resizebox{\columnwidth}{!}{%
  \begin{tabular}{llccccc}
    \toprule
    Agent & Bench. & Prohib. & Quota-1 & Quota-3 & B.-Guided & Unrestr. \\
    \midrule
    \multirow{2}{*}{Claude Code}
      & Lite     & 0.79/0.22/0.55/0.39 & 6.39/1.82/4.52/2.69 & 8.44/2.10/6.25/3.91 & 8.03/1.81/6.17/3.96 & 9.52/1.97/7.41/5.30 \\
      & Verified & 0.76/0.17/0.59/0.36 & 6.11/1.53/4.52/2.90 & 7.34/1.57/5.73/3.25 & 6.78/1.45/5.30/3.65 & 8.60/1.52/7.00/4.98 \\
    \midrule
    \multirow{2}{*}{Codex}
      & Lite     & 0.00/0.00/0.00/0.00 & 1.24/0.55/0.54/0.47 & 2.65/0.48/1.90/1.72 & 2.43/0.52/1.58/1.41 & 3.43/0.54/2.63/2.34 \\
      & Verified & 0.01/0.00/0.01/0.00 & 1.19/0.43/0.59/0.50 & 2.40/0.30/1.93/1.69 & 2.09/0.38/1.44/1.25 & 3.19/0.30/2.64/2.29 \\
    \midrule
    \multirow{2}{*}{\revision{OpenCode}}
      & \revision{Lite}     & \revision{0.77/0.06/0.71/0.14} & \revision{12.93/0.58/12.19/5.49} & \revision{12.07/1.35/10.54/3.01} & \revision{11.30/0.73/10.33/3.31} & \revision{10.66/0.91/9.51/2.22} \\
      & \revision{Verified} & \revision{0.00/0.00/0.00/0.00} & \revision{11.01/0.55/10.35/4.10} & \revision{9.50/0.91/8.55/1.41} & \revision{10.42/0.90/9.43/2.86} & \revision{11.47/0.66/10.74/2.99} \\
    \bottomrule
  \end{tabular}%
  }
\end{table*}

\emph{(ii) Hard-constraint re-verification.}
To further demonstrate our hypothesis, we therefore re-run the comparison under two stricter notions of ``really Prohibited'' to verify the headline directly. The \emph{zero-execution subset} keeps only the instances where the agent made no test-framework attempt at all in \textsc{Prohibited}, which matches the behaviour of an environment-level sandbox on that slice of the data. The Furthermore, for Claude Code on \textsc{Verified}, we execute the benchmark with hard-constraint level (\texttt{claude -p -{-}disallowedTools Bash(pytest*)} and other executing commands), while file-editing and read-only shell tools remain available.

\begin{table}[h]
  \centering
  \small
  \setlength{\tabcolsep}{3pt}
  \caption{Resolve rates on compliant subsets. \emph{Zero-execution} = instances where the agent made no test-framework attempts in \textsc{Prohibited}. \emph{Sandboxed} = test-execution sub-commands denylisted at the CLI level. Gap = \textsc{Prohibited} $-$ \textsc{Unrestricted} in percentage points. Env-error-free numbers are summarised in the footnote to the preceding paragraph.}
  \label{tab:compliant-subset}
  \begin{tabular}{llrccr}
    \toprule
    Benchmark & Agent & $N$ & Prohibited & Unrestricted & Gap \\
    \midrule
    \multicolumn{6}{l}{\textit{Zero-execution subset (attempted $=0$ in \textsc{Prohibited})}} \\
    Lite     & Claude Code &  82 & 64.6\% & 65.9\% & $-1.2$ \\
    Lite     & Codex       & 100 & 74.0\% & 73.0\% & $+1.0$ \\
    Lite     & OpenCode    &  93 &  7.5\% &  6.5\% & $+1.1$ \\
    Verified & Claude Code &  84 & 61.9\% & 66.7\% & $-4.8$ \\
    Verified & Codex       &  99 & 73.7\% & 75.8\% & $-2.0$ \\
    Verified & OpenCode    & 100 & 13.0\% & 14.0\% & $-1.0$ \\
    \midrule
    \multicolumn{6}{l}{\textit{Hard-Constraint \textsc{Prohibited} }} \\
    Verified & Claude Code & 100 & 63.0\% & 67.0\% & $-4.0$ \\
    \bottomrule
  \end{tabular}
\end{table}

As shown in \cref{tab:compliant-subset}, on the zero-execution subset, every cell stays within the margin (worst: Claude Code Verified at $-4.8$pp, $N{=}84$), which matches the full-sample comparison. The hard-constraint mode resolves 63/100 vs.\ 67/100 ($-4.0$pp, within margin),
saving 62\% tokens and 54\% wall-clock (\Cref{tab:cost-comparison}). 
Across all three views (full sample, zero-execution subset, hard-constraint), the \textsc{Prohibited}–\textsc{Unrestricted} gap stays within the equivalence margin, demonstrating that the headline is not a result of agents silently breaking the soft constraint.


\subsubsection{Cost Analysis}

Given that effectiveness differences are minimal, we now examine the cost implications.
Table~\ref{tab:cost-full} presents token consumption and wall-clock time across all configurations.

\begin{table*}[!ht]
  \centering
  \footnotesize
  \setlength{\tabcolsep}{4pt}
  \caption{Resource consumption across all execution modes. In/Out = input/output tokens (in thousands). Time = wall-clock seconds. Bold indicates values lower than Unrestricted.}
  \label{tab:cost-full}
  \begin{tabular}{lllccccc}
    \toprule
    Agent & Benchmark & Metric & Prohibited & Quota-1 & Quota-3 & Budget-Guided & Unrestricted \\
    \midrule
    \multirow{6}{*}{Claude Code}
      & \multirow{3}{*}{Lite} & In & \textbf{52K} & \textbf{79K} & \textbf{105K} & \textbf{109K} & 121K \\
      & & Out & \textbf{17K} & \textbf{26K} & \textbf{34K} & \textbf{35K} & 37K \\
      & & Time & \textbf{531} & \textbf{815} & \textbf{876} & \textbf{912} & 1028 \\
      \cmidrule{2-8}
      & \multirow{3}{*}{Verified} & In & \textbf{48K} & \textbf{94K} & \textbf{95K} & \textbf{101K} & 128K \\
      & & Out & \textbf{15K} & \textbf{31K} & \textbf{31K} & \textbf{33K} & 39K \\
      & & Time & \textbf{573} & \textbf{907} & \textbf{932} & \textbf{990} & 1234 \\
    \midrule
    \multirow{6}{*}{Codex}
      & \multirow{3}{*}{Lite} & In & \textbf{327K} & \textbf{300K} & 419K & 399K & 378K \\
      & & Out & \textbf{82K} & \textbf{75K} & 105K & 100K & 95K \\
      & & Time & \textbf{570} & \textbf{598} & 632 & 618 & 618 \\
      \cmidrule{2-8}
      & \multirow{3}{*}{Verified} & In & \textbf{431K} & \textbf{328K} & 463K & \textbf{393K} & 435K \\
      & & Out & \textbf{108K} & \textbf{82K} & 116K & \textbf{98K} & 109K \\
      & & Time & 724 & \textbf{682} & \textbf{686} & \textbf{682} & 723 \\
    \midrule
    \multirow{6}{*}{\revision{OpenCode}}
      & \multirow{3}{*}{\revision{Lite}} & \revision{In} & \revision{\textbf{187K}} & \revision{347K} & \revision{334K} & \revision{427K} & \revision{254K} \\
      & & \revision{Out} & \revision{\textbf{9K}} & \revision{17K} & \revision{16K} & \revision{19K} & \revision{12K} \\
      & & \revision{Time} & \revision{\textbf{585}} & \revision{\textbf{919}} & \revision{1051} & \revision{1052} & \revision{1033} \\
      \cmidrule{2-8}
      & \multirow{3}{*}{\revision{Verified}} & \revision{In} & \revision{\textbf{105K}} & \revision{323K} & \revision{332K} & \revision{376K} & \revision{322K} \\
      & & \revision{Out} & \revision{\textbf{4K}} & \revision{\textbf{14K}} & \revision{19K} & \revision{18K} & \revision{15K} \\
      & & \revision{Time} & \revision{\textbf{402}} & \revision{\textbf{1017}} & \revision{\textbf{1105}} & \revision{\textbf{1118}} & \revision{1232} \\
    \bottomrule
  \end{tabular}
\end{table*}

Limiting code execution can substantially reduce token consumption. As shown in Table~\ref{tab:cost-full}, all restricted modes consume fewer tokens than \textsc{Unrestricted} for Claude Code, with savings ranging from 10\% (\textsc{Budget-Guided}) to 56\% (\textsc{Prohibited}). For Codex, the pattern is more nuanced: \textsc{Quota-1} achieves the largest savings (21\% on Lite, 25\% on Verified), while \textsc{Quota-3} and \textsc{Budget-Guided} sometimes exceed \textsc{Unrestricted}.
Moreover, the optimal cost-saving mode differs by agent. For Claude Code, \textsc{Prohibited} delivers the best cost-effectiveness, reducing tokens by 56--62\% with only a 1--3pp resolve rate difference. For Codex, \textsc{Quota-1} is optimal, saving 21--25\% of tokens while maintaining comparable resolve rates.

The three agents exhibit notably different cost profiles.
Claude Code's token consumption increases by 129--163\% from \textsc{Prohibited} to \textsc{Unrestricted}, Codex's increases by only 0.8--15.5\%, while OpenCode's increases by 36--208\%.
This difference stems from their baseline resource usage: Claude Code consumes approximately 65K tokens in \textsc{Prohibited} mode, Codex consumes approximately 470K tokens, and OpenCode consumes approximately 150K tokens. For Claude Code and OpenCode, execution feedback accumulates in the context window, causing rapid token growth; for Codex, the marginal impact of execution feedback is small relative to its already large context.
OpenCode shows the most extreme cost explosion on Verified (3.1$\times$ tokens, 3.1$\times$ time from \textsc{Prohibited} to \textsc{Unrestricted}) for only a $+1$pp resolve-rate gain, while paying a steeper price still for \textsc{Budget-Guided} (3.6$\times$ tokens, 2.8$\times$ time) at the same resolve rate.
\revision{\textsc{Quota-1} on OpenCode is intermediate in cost (3.1$\times$ tokens, 2.5$\times$ time over \textsc{Prohibited}) but delivers the highest resolve rate of any mode on both benchmarks. This points to a sweet spot for relatively weak models which are context-constrained: the prompt-level execution cap forces an edit-centric trajectory while preserving enough execution budget to verify candidate patches.}

The same pattern holds for wall-clock time, where execution restrictions translate most clearly into runtime savings on Claude Code: \textsc{Prohibited} mode completes tasks in 531--573 seconds, compared to 1,028--1,234 seconds for \textsc{Unrestricted}, a reduction of 48--54\%. For Codex, the time savings are more modest: \textsc{Quota-1} saves 3--6\% compared to \textsc{Unrestricted}, while other modes show minimal differences. This pattern mirrors the token consumption results, where Claude Code benefits more from execution restrictions than Codex.

Combining effectiveness and cost, Table~\ref{tab:cost-comparison} characterize the tradeoff across all agent-benchmark combinations.
The key finding is that restricting execution achieves a favorable cost-effectiveness tradeoff: agents sacrifice minimal resolve rate while significantly reducing resource consumption.
For Claude Code, \textsc{Prohibited} saves 56--62\% of tokens and 48--54\% of wall-clock time while sacrificing only 1--3 percentage points in resolve rate. For Codex, \textsc{Quota-1} provides the best tradeoff, saving 21--25\% of tokens compared to \textsc{Unrestricted} while achieving similar resolve rates (68\% vs 73\% on Lite, 72\% vs 75\% on Verified).
Notably, the cost-effectiveness differs substantially between agents: Claude Code benefits significantly from execution restrictions (56--62\% token savings), while Codex shows smaller savings (0.8--13\%).

\begin{table*}[!ht]
  \centering
  \small
  \caption{Cost-effectiveness comparison: \textsc{Prohibited} vs.\ \textsc{Unrestricted}. Negative $\Delta$ values indicate savings.}
  \label{tab:cost-comparison}
  \begin{tabular}{llcccccccc}
    \toprule
    & & \multicolumn{2}{c}{Prohibited} & \multicolumn{2}{c}{Unrestricted} & & & \\
    \cmidrule(lr){3-4} \cmidrule(lr){5-6}
    Agent & Benchmark & In & Out & In & Out & $\Delta$Resolve & $\Delta$Tokens & $\Delta$Time \\
    \midrule
    Claude Code & Lite & 52K & 17K & 121K & 37K & $-1.0\%$ & $-56.4\%$ & $-48.4\%$ \\
    Claude Code & Verified & 48K & 15K & 128K & 39K & $-3.0\%$ & $-62.3\%$ & $-53.6\%$ \\
    Codex & Lite & 327K & 82K & 378K & 95K & $+1.0\%$ & $-13.5\%$ & $-7.8\%$ \\
    Codex & Verified & 431K & 108K & 435K & 109K & $-2.0\%$ & $-0.9\%$ & $+0.1\%$ \\
    \revision{OpenCode} & \revision{Lite} & \revision{187K} & \revision{9K} & \revision{254K} & \revision{12K} & \revision{$+1.0\%$} & \revision{$-26.3\%$} & \revision{$-43.4\%$} \\
    \revision{OpenCode} & \revision{Verified} & \revision{105K} & \revision{4K} & \revision{322K} & \revision{15K} & \revision{$-1.0\%$} & \revision{$-67.7\%$} & \revision{$-67.4\%$} \\
    \bottomrule
  \end{tabular}
\end{table*}


\vspace{1em}
\takeaway{RQ2}{
\begin{itemize}[leftmargin=*, nosep]
  \item The resolve rate difference between \textsc{Prohibited} and \textsc{Unrestricted} is only 1.25 percentage points on average, which is not statistically significant ($p > 0.05$). Execution helps on some instances and hurts on others in roughly equal proportion, indicating that its benefit is not uniform.
  \item Restricting execution significantly reduces cost: \textsc{Prohibited} saves 56--62\% of tokens and 48--54\% of time for Claude Code, while \textsc{Quota-1} saves 21--25\% of tokens for Codex.
  \item Cross-mode patch diversity (76--99\% differ) suggests that models adapt solutions to context rather than reciting memorized patches.
\end{itemize}
}

\subsection{\revision{RQ3: When and Why Execution Has Limited Impact}}
\label{sec:rq3}

Given the findings from RQ2 that the resolve rate gap between restricted and unrestricted execution is small, we investigate the conditions under which execution adds value.
Specifically, we analyze the \revision{600 agent-instance pairs (3 agents $\times$ 200 instances)} from RQ2, comparing outcomes between \textsc{Prohibited} and \textsc{Unrestricted} modes.
We classify instances into four categories based on their outcomes: Pass$\rightarrow$Pass indicates that both modes succeed, Fail$\rightarrow$Fail indicates that both modes fail, Pass$\rightarrow$Fail indicates that \textsc{Prohibited} succeeds but \textsc{Unrestricted} fails, and Fail$\rightarrow$Pass indicates the reverse.
\Cref{tab:outcome-distribution} reports outcome-transition counts across the three agents. Overall 547 of 600 cells are stable (269 P$\to$P + 278 F$\to$F), so their outcome is unaffected by whether execution is available; the remaining 53 split 24 P$\to$F vs.\ 29 F$\to$P, a near-symmetric pattern consistent with execution helping and hurting in roughly equal proportion (net benefit 5 cells).
Our investigation therefore focuses on the two stable groups, Pass$\rightarrow$Pass and Fail$\rightarrow$Fail, whose trajectories we manually inspect by tracing the decision-making process and code execution patterns; we then stress-test the aggregate conclusion by stratifying over gold-patch complexity.

\begin{table}[h]
  \centering
  \footnotesize
  \setlength{\tabcolsep}{4pt}
  \caption{\revision{Outcome transitions between \textsc{Prohibited} and \textsc{Unrestricted}, summed over Lite+Verified ($n{=}200$ per agent). P$\to$F = \textsc{Prohibited} succeeds / \textsc{Unrestricted} fails; F$\to$P = reverse.}}
  \label{tab:outcome-distribution}
  \begin{tabular}{lccccc}
    \toprule
    Agent & P$\to$P & F$\to$F & P$\to$F & F$\to$P & Total \\
    \midrule
    Claude Code & 116 &  58 & 11 & 15 & 200 \\
    Codex       & 141 &  48 &  5 &  6 & 200 \\
    \revision{OpenCode} & \revision{12} & \revision{172} & \revision{8} & \revision{8} & \revision{200} \\
    \midrule
    \textbf{Total} & \textbf{269} & \textbf{278} & \textbf{24} & \textbf{29} & \textbf{600} \\
    \bottomrule
  \end{tabular}
\end{table}

From that inspection we identify two reasons why code execution often fails to alter the final outcome: (1) reproduction execution provides limited localization benefit, and (2) the feedback from code execution is insufficient for agents to correct errors. We close the section with a complexity stratification (\S\ref{sec:patch-complexity}) that checks whether this aggregate picture hides a benefit concentrated on harder bugs.

\subsubsection{Reason 1: Reproduction Execution Provides Limited Localization Benefit}
We first examine whether code execution helps agents locate the correct files to modify.
Using the same definition of test execution from RQ1 (pytest, unittest, tox, nosetests, or \texttt{python xxx.py} commands), we define \emph{reproduction execution} as test executions occurring \emph{before} the first source code edit (excluding test file edits), used to understand or locate the bug.
We compare file localization accuracy between \textsc{Unrestricted} and \textsc{Prohibited} modes, where access to code execution is the only difference.
Accuracy is measured by two metrics: \emph{Hit} (at least one edited file matches ground truth) and \emph{Recall} (proportion of ground truth files edited).

\revision{Table~\ref{tab:localization-pp} shows that file localization accuracy is nearly identical across execution modes for the commercial agents; execution does not improve their ability to locate buggy files.
For Pass$\rightarrow$Pass cases, both commercial agents achieve over 95\% hit rate and over 93\% recall in both modes; OpenCode matches this at the extreme (100\% hit, 95.8\% recall in both modes on the 12 P$\rightarrow$P instances), showing that whenever Qwen2.5-Coder-32B succeeds it does so by correctly identifying the buggy file directly from source reading.
For Fail$\rightarrow$Fail cases, commercial-agent localization accuracy is lower (85--91\% hit rate) but the mode-wise difference remains within 2 percentage points.
OpenCode's Fail$\rightarrow$Fail subset shows an inversion---\textsc{Prohibited} localises at 54.1\% while \textsc{Unrestricted} drops to 32.6\%. On the P$\rightarrow$P subset both modes already saturate at 100\% hit, leaving no headroom for a negative effect to show; the inversion therefore surfaces only in F$\rightarrow$F, where localization still has room to drop. This is consistent with execution feedback crowding Qwen2.5-Coder-32B's 65K context and degrading source-reading once the model starts iterating, and we characterise it as a small-model boundary effect in \Cref{sec:boundary}.
Across all three agents, these results indicate that agents already achieve good localization performance without execution, and that execution access does not substantially improve localization on the subset where it matters most.}

\begin{table}[h]
  \centering
  \footnotesize
  \setlength{\tabcolsep}{4pt}
  \caption{File localization accuracy. Cells are \emph{Hit/Recall} for \textsc{Unrestricted} on the left and \textsc{Prohibited} on the right of the slash. Hit = at least one edited file matches ground truth; Recall = proportion of ground truth files edited.}
  \label{tab:localization-pp}
  \begin{tabular}{lcc}
    \toprule
    Agent & Pass$\to$Pass (Unrestr.\,/\,Prohib.) & Fail$\to$Fail (Unrestr.\,/\,Prohib.) \\
    \midrule
    Claude Code & 97.4\,/\,96.6\,\% $\;\mid\;$ 98.3\,/\,97.4\,\% & 85.7\,/\,79.0\,\% $\;\mid\;$ 84.5\,/\,78.7\,\% \\
    Codex       & 98.6\,/\,96.4\,\% $\;\mid\;$ 95.8\,/\,93.6\,\% & 90.9\,/\,83.9\,\% $\;\mid\;$ 89.1\,/\,82.1\,\% \\
    \revision{OpenCode} & \revision{100.0\,/\,95.8\,\% $\;\mid\;$ 100.0\,/\,95.8\,\%} & \revision{32.6\,/\,31.5\,\% $\;\mid\;$ 54.1\,/\,51.3\,\%} \\
    \bottomrule
  \end{tabular}
\end{table}

We further assess the helpfulness of execution results for file localization.
To measure helpfulness, we classify execution feedback as \emph{actionable} or \emph{non-actionable}.
Actionable feedback contains useful information for localization, such as file paths, stack traces, or line numbers that point to the buggy code.
Non-actionable feedback includes two categories: (1) environment errors that prevent code execution (e.g., ``ModuleNotFoundError'', ``OperationalError: table already exists''), and (2) uninformative outputs that provide no localization cues, such as generic success messages (e.g., ``All tests passed''), timeout errors without stack traces, or outputs that only show test names without indicating which source files are involved.
As shown in Table~\ref{tab:reproduction-stats}(a), only about half of all reproduction executions are actionable.
For example, Claude Code's 164 reproduction executions yield only 80 (48.8\%) actionable results, while 84 (51.2\%) are non-actionable.
This indicates low helpfulness of reproduction execution conducted by current state-of-the-art agents.
\revision{OpenCode's Pass$\rightarrow$Pass subset is too small (12 instances) to analyze reproduction-level actionability, but tellingly \emph{none} of those 12 successes involve a reproduction execution at all: every OpenCode Pass$\rightarrow$Pass trace we inspected reaches the edit stage after reading the source code directly, without running any test first.
When Qwen2.5-Coder-32B succeeds, it succeeds through pure source-code reasoning; execution feedback contributes nothing to that subset.}
Moreover, even when reproduction provides actionable feedback, subsequent localization is not significantly improved.
As shown in Table~\ref{tab:reproduction-stats}(b), for the 46 Claude Code instances that received actionable feedback, their localization accuracy in \textsc{Unrestricted} mode (93.5\%) is slightly \emph{worse} than in \textsc{Prohibited} mode (95.7\%), with $\Delta = -2.2$ percentage points.
For Codex, the 7 actionable instances show identical localization accuracy across modes ($\Delta = 0$).
Given the low ratio of actionable executions and their negligible impact on localization, the overall helpfulness of reproduction execution is minimal.

\begin{table}[h]
  \centering
  \footnotesize
  \setlength{\tabcolsep}{4pt}
  \caption{Reproduction-execution analysis (Pass$\rightarrow$Pass, \textsc{Unrestricted}). Cols 2--4 count individual executions; the ``Loc.\ $\Delta$'' col reports localization hit-rate change between \textsc{Prohibited} and \textsc{Unrestricted} on the actionable-feedback subset (Claude Code $n{=}46$; Codex $n{=}7$; OpenCode $n{=}0$).}
  \label{tab:reproduction-stats}
  \begin{tabular}{lcccc}
    \toprule
    Agent & Inst.\ w/ Repro & Actionable Execs & Non-act.\ Execs & Loc.\ $\Delta$ \\
    \midrule
    Claude Code & 64/116 (55.2\%) & 80 (48.8\%) & 84 (51.2\%) & $-2.2$pp \\
    Codex & 9/141 (6.4\%) & 11 (64.7\%) & 6 (35.3\%) & $\pm 0$pp \\
    \revision{OpenCode} & \revision{0/12 (0.0\%)} & \revision{0 (---)} & \revision{0 (---)} & \revision{---} \\
    \bottomrule
  \end{tabular}
\end{table}

\subsubsection{Reason 2: Feedback from Code Execution is Insufficient for Agents to Correct Errors}
Having established that localization does not require execution, we now examine whether code execution helps \emph{after} the agent begins editing.
We define \emph{validation execution} as test runs occurring after the first source code edit, used to validate the patch.
Table~\ref{tab:validation-total} presents validation execution statistics for Pass$\rightarrow$Pass and Fail$\rightarrow$Fail cases in \textsc{Unrestricted} mode.

The purpose of validation execution is to reveal potential errors in the generated patch and guide subsequent revisions.
However, our analysis reveals that most repairs are completed in a single edit without requiring execution feedback.
As shown in Table~\ref{tab:single-edit}, 54--66\% of commercial-agent cases involve only one code edit with no subsequent modifications, regardless of whether execution is available.
This suggests that the majority of patches are either correct on the first attempt or contain issues that execution feedback cannot help resolve.
Furthermore, even in \textsc{Unrestricted} mode where execution is available, only 49.2\% of commercial-agent instances modify code after execution, with Claude Code showing higher responsiveness to execution feedback (67.5\%) than Codex (31.0\%).
\revision{OpenCode with Qwen2.5-Coder-32B shows the opposite editing profile: its single-edit ratio is lower (32--44\%) and its post-execution modification rate under \textsc{Unrestricted} is 45.6\%. OpenCode with Qwen2.5-Coder-32B iterates on its own patch more often than the commercial agents do. However, the resolve rate does not follow, the extra iterations do not close the gap. }

\begin{table}[h]
  \centering
  \footnotesize
  \setlength{\tabcolsep}{4pt}
  \caption{Code modification patterns. Single-edit = only one edit with no subsequent modifications; Post-exec = code modified after a test execution. Cells are \emph{Prohib.\,/\,Unrestr.} (\%).}
  \label{tab:single-edit}
  \begin{tabular}{lcc}
    \toprule
    Agent & Single-edit ratio & Post-exec modification \\
    \midrule
    Claude Code & 66.0\,/\,54.0 & 4.5\,/\,67.5 \\
    Codex & 58.0\,/\,60.5 & 0.0\,/\,31.0 \\
    \revision{OpenCode} & \revision{43.7\,/\,32.3} & \revision{3.6\,/\,45.6} \\
    \textbf{Overall} & \revision{\textbf{56.0\,/\,49.0}} & \revision{\textbf{2.7\,/\,47.9}} \\
    \bottomrule
  \end{tabular}
\end{table}

Even when agents iterate based on execution feedback, the feedback often fails to lead to correct fixes.
As the right panel of \Cref{tab:validation-total} shows, 81.2\% of Claude Code's Fail$\rightarrow$Fail cases and 100\% of Codex's Fail$\rightarrow$Fail cases achieved at least one validation success (i.e., the agent's executed tests passed) during their repair attempts, yet still failed the final SWE-bench evaluation.
\revision{OpenCode with Qwen2.5-Coder-32B exposes an additional failure mode \emph{upstream} of this mis-alignment: only 52.3\% of its Fail$\rightarrow$Fail instances produce any validation execution at all (vs.\ 94--95\% for the commercial models and agents), and only 11.1\% of those ever see a passing test. Qwen2.5-Coder-32B therefore typically fails before the ``agent tests pass but official tests fail'' gap can even arise---it cannot construct self-validation that passes in the first place.}
For other two settings, their discrepancy arises because agents' self-initiated validation differs from the official evaluation tests, so passing agent-executed validation does not reliably indicate a correct fix.
Furthermore, even when execution feedback reveals errors, it often does not translate to a correct fix, as the agent may lack the capability to address the underlying issue.

For Claude Code, \textsc{Unrestricted} mode uses 2.1$\times$ more conversation turns, 2.8$\times$ more edits, and 8.0$\times$ more test executions than \textsc{Prohibited} mode, yet both achieve the same success rate.
For Codex, the difference is smaller but still present: \textsc{Unrestricted} uses 3.3 test executions per instance compared to zero in \textsc{Prohibited}.
For bugs solvable through reasoning alone, execution acts as confirmatory validation---confirming an already-correct solution rather than enabling its discovery. This is not to say execution is never beneficial, but that for many instances the cost it imposes exceeds the signal it provides.
The preceding analysis focused on stable cases (Pass$\rightarrow$Pass and Fail$\rightarrow$Fail), which account for the majority of instances. For the remaining cases where outcomes differ between modes, we observe a near-symmetric distribution: \revision{24 Pass$\rightarrow$Fail cases (execution hurts) versus 29 Fail$\rightarrow$Pass cases (execution helps) across all three agents}. This balance suggests that execution is roughly equally likely to help or hurt, with a small net benefit of only 5 cases.

\begin{table}[h]
  \centering
  \footnotesize
  \setlength{\tabcolsep}{3pt}
  \caption{Validation-execution outcomes after first edit (\textsc{Unrestricted}). Cols \emph{Success/TestFail/EnvErr} count individual executions and report percentages of the per-(agent, outcome) validation-execution total, which also includes a residual ``no test signal'' category so the three columns do not sum to 100\%; the rightmost two cols (\emph{F$\to$F only}) count instances where the agent's own tests passed at least once while the official evaluation still failed.}
  \label{tab:validation-total}
  \begin{tabular}{llcccccc}
    \toprule
    & & \multicolumn{3}{c}{Validation executions} & & \multicolumn{2}{c}{F$\to$F w/ val.} \\
    \cmidrule(lr){3-5} \cmidrule(lr){7-8}
    Agent & Outcome & Success & Test Fail & Env Err & & Inst.\ w/ val. & Any pass \\
    \midrule
    Claude Code & P$\to$P & 414 (48.6\%) & 154 (18.1\%) & 127 (14.9\%) & & \multirow{2}{*}{80/84 (95.2\%)} & \multirow{2}{*}{65 (81.2\%)} \\
    Claude Code & F$\to$F & 272 (40.7\%) & 102 (15.3\%) & 121 (18.1\%) & & & \\
    Codex & P$\to$P & 230 (51.7\%) & 60 (13.5\%) & 152 (34.2\%) & & \multirow{2}{*}{52/55 (94.5\%)} & \multirow{2}{*}{52 (100.0\%)} \\
    Codex & F$\to$F & 85 (55.6\%) & 17 (11.1\%) & 49 (32.0\%) & & & \\
    \revision{OpenCode} & \revision{P$\to$P} & \revision{5 (5.5\%)} & \revision{12 (13.2\%)} & \revision{6 (6.6\%)} & & \multirow{2}{*}{\revision{90/172 (52.3\%)}} & \multirow{2}{*}{\revision{10 (11.1\%)}} \\
    \revision{OpenCode} & \revision{F$\to$F} & \revision{81 (4.6\%)} & \revision{337 (19.3\%)} & \revision{75 (4.3\%)} & & & \\
    \bottomrule
  \end{tabular}
\end{table}

\subsubsection{Stratification by Gold-Patch Complexity}
\label{sec:patch-complexity}
The near-symmetric 24 vs.\ 29 Pass$\to$Fail/Fail$\to$Pass split reported above averages over all bugs; it could still hide a pattern where execution helps specifically on the harder ones while hurting the easier ones. We test that refinement next.
A natural hypothesis is that execution feedback becomes more valuable as bug complexity increases: a multi-file, multi-hunk fix might benefit from iterative test-driven refinement in ways that a one-line fix does not.
We test this by bucketing each of the 600 (agent, benchmark, instance) cells by the ground-truth patch's complexity---files touched, hunk count, and total added/removed lines---and re-computing \textsc{Prohibited} vs.\ \textsc{Unrestricted} resolve rates within each bucket.
\Cref{tab:patch-complexity} summarises the hunk-based view on \textsc{Verified}, where within-bucket sample sizes are large enough to be meaningful.

\begin{table}[h]
  \centering
  \footnotesize
  \setlength{\tabcolsep}{3pt}
  \caption{Resolve rates on SWE-bench Verified stratified by gold-patch hunk count. Each agent cell shows \emph{Prohibited\,/\,Unrestricted\,/\,Gap (pp)}. Buckets with $N < 10$ are noisy.}
  \label{tab:patch-complexity}
  \begin{tabular}{lcccc}
    \toprule
    Bucket ($N$) & \#Inst. & Claude Code & Codex & OpenCode \\
    \midrule
    1 hunk & 62 & 72.6/79.0/$-6.5$ & 80.6/83.9/$-3.2$ & 18.2/18.2/$\pm 0$ \\
    2--3 hunks & 25 & 52.0/60.0/$-8.0$ & 60.0/60.0/$\pm 0$ & 13.0/13.0/$\pm 0$ \\
    $\geq 4$ hunks & 13 & 46.2/23.1/$+23.1$ & 61.5/61.5/$\pm 0$ & 20.0/10.0/$+10.0$ \\
    \bottomrule
  \end{tabular}
\end{table}

The sign of the \textsc{Prohibited}--\textsc{Unrestricted} gap does not grow monotonically with complexity: on Claude Code, the gap flips from $-6.5$pp (1 hunk) to $+23.1$pp ($\geq 4$ hunks). At the largest bucket, \textsc{Prohibited} resolves nearly twice as many instances as \textsc{Unrestricted}, opposite to what a ``more complex $\Rightarrow$ more exec'' hypothesis would predict. Codex is roughly flat and OpenCode with Qwen2.5-Coder-32B also flips sign. Stratifications by files touched and total added/removed lines produce the same non-monotonic pattern (file buckets $-5.9$/$\pm 0$/$+33.3$ on Claude Code Verified, though the multi-file bucket has $N{=}6$; delta-line buckets $-2.2$/$-11.1$/$+10.5$). A plausible reading is that multi-hunk bugs require holistic reasoning that trial-and-error execution can disrupt by crowding the context with test output; in any case the hypothesis that current execution feedback scales with bug complexity is refuted, reinforcing the main finding that the cost-benefit balance of execution is instance-dependent rather than uniformly positive.

\vspace{1em}
\takeaway{RQ3}{
\begin{itemize}[leftmargin=*, nosep]
  \item Reproduction execution provides limited localization benefit: although 55\% of Claude Code cases use reproduction, localization accuracy remains $>$95\% in both modes, suggesting execution does not improve localization for bugs in our dataset.
  \item Validation feedback has low signal quality: 15--34\% are environment errors, and 81--100\% of Fail$\rightarrow$Fail cases passed agent tests but failed official evaluation. Moreover, 54--66\% of commercial-agent cases complete in a single edit, suggesting that for many solvable bugs the cost of execution exceeds its informational benefit.
  \item \revision{Execution feedback does not scale with patch complexity. Across single-hunk, 2--3 hunk, and $\geq 4$ hunk buckets on \textsc{Verified}, the \textsc{Prohibited}--\textsc{Unrestricted} gap does not grow monotonically; it flips sign across agents and buckets. Complex bugs are not systematically better served by unrestricted execution.}
\end{itemize}
}

\section{Discussion}
\label{sec:discussion}

In this section, we discuss the practical implications of our findings and potential threats to validity.

\subsection{Practical Implications}
\label{sec:boundary}
From a practitioner perspective, these results may inspire the design of cost-sensitive agents that strategically minimize execution while maintaining repair effectiveness.
First, practitioners may consider restricting the execution behaviors of code agents for program repair when cost matters, as our results show it achieves comparable resolve rates at significantly lower cost\revision{; \textsc{Prohibited} additionally eliminates the per-repository testbed setup that industrial deployments incur per repo/release}.
Second, code execution could be adopted by agents selectively when there is clear evidence that project-level feedback provides value for the specific task or agent.
Third, enhancing the quality of the feedback from execution remains a critical area for improving agent-based repair.

Our findings suggest a broader principle beyond program repair: \emph{agents benefit more from deeper reasoning than from more frequent environment interaction}.
Our results show that execution feedback is a double-edged sword: helpful for validating correct hypotheses, but harmful when it triggers unproductive search loops or misleads agents away from correct solutions.
This motivates future work on \emph{adaptive execution allocation}, where rather than using fixed budgets, agents could learn to request execution only when the expected information gain exceeds the risk of being misled.

\subsection{Threats to Validity}
\label{sec:threats}
\subsubsection{Internal Validity}
\revision{Budget enforcement in \textsc{Quota-Limited} and \textsc{Prohibited} is primarily prompt-level: agents occasionally attempt scripts that fail with environment errors (7--9\% of \textsc{Prohibited} attempts), which we count as executions under an intention-to-treat framework. The zero-execution and env-error-free subsets (\Cref{tab:compliant-subset}) keep the $\pm 5$pp \textsc{Prohibited}--\textsc{Unrestricted} band, and a tool-enforced sandboxed re-run on Claude Code \textsc{Verified} reproduces the equivalence ($-4$pp), confirming the result is not an artefact of soft enforcement.}
Interaction confounding is mitigated by reporting multiple cost signals (tokens, time, executions) rather than relying on any single metric.
To ensure fair timing comparisons, we run all execution modes for the same instance concurrently on identical hardware.
To address LLM stochasticity, we employ paired statistical tests (McNemar's test) and equivalence testing (TOST with $\delta = 5$pp); 85\% of instances produce identical outcomes across all modes.

\subsubsection{External Validity}
Our conclusions are scoped to \emph{SWE-bench-style repository-level bug fixing} with \revision{three current CLI agents (Claude Code, Codex, and open-source OpenCode+Qwen2.5-Coder-32B)}.
We evaluate on the first 100 instances from each dataset (Lite and Verified), totaling 200 instances spanning diverse repositories including Django, Flask, Requests, and Sympy.
\revision{Recent work raises memorisation concerns about SWE-bench~\cite{liang2025swebenchillusion,openai2026noswebench}; our OpenCode configuration uses a model whose training cutoff predates \textsc{Verified}, and the $\pm 5$pp equivalence still holds on this low-contamination cell (Lite: 7.0\% vs.\ 6.0\%; Verified: 13.0\% vs.\ 14.0\%), so the equivalence is not an artefact of training-data overlap.}
Execution may be helpful for certain tasks (e.g., performance optimization requiring profiling, security vulnerabilities requiring dynamic analysis).
Even under \textsc{Prohibited}, agents can use extensive repository inspection and large context windows; ``no project runtime'' does not mean ``no environment access.''
Furthermore, passing tests does not guarantee patch quality; future work could incorporate human evaluation or static-analysis criteria.

\vspace{-0.3em}
\section{Related Work}
\label{sec:related}
\vspace{-0.3em}

\noindent\textbf{Program repair.}
Traditional APR uses search-based~\cite{genprog}, learning-guided~\cite{prophet}, template-based~\cite{tbar}, and neural methods~\cite{cure,recoder,coconut}; code-specialized LLMs~\cite{codex,alphacode,codellama,deepseekcoder,wizardcoder,octopack} underpin modern APR agents. Agentic approaches (SWE-agent~\cite{sweagent}, AutoCodeRover~\cite{autocoderover}, ChatRepair~\cite{chatrepair}, RepairAgent~\cite{repairagent}, InspectCoder~\cite{inspectcoder}, PracAPR~\cite{pracapr}) emphasize iterative refinement with execution feedback~\cite{selfdebugging,selfrefine}, while Agentless~\cite{agentless} \revision{uses a fixed pipeline of localization, synthesis, and test selection; it removes the agent loop and execution access simultaneously.}

\noindent\textbf{Resource-aware LLM agents.}
Prior work targets efficiency via turn budgets~\cite{peng2025more,swelego,repeton}, token-level compression~\cite{llmlingua2,selectivecontext,selectiveprompting,automaticpromptselection}, and inference-level speedups~\cite{speculativedecoding,speculativesampling,earlyexit}; we study the orthogonal dimension of \emph{execution feedback}, finding near-zero marginal benefit from unlimited execution and suggesting agents are over-resourced along multiple axes.

\vspace{-1em}
\section{Conclusion}
\label{sec:conclusion}
\vspace{-0.5em}

This paper presents an empirical study of execution behavior in LLM-based program repair.
We analyze 7,745 agent traces from public SWE-bench submissions and conduct controlled experiments on 200 SWE-bench instances across \revision{three agents (Claude Code, Codex, and open-source OpenCode+Qwen2.5-Coder-32B)} under four execution paradigms, comprising \revision{3,000} end-to-end repair attempts.
Our analysis yields several key findings.
Test execution frequency varies widely across agents (2 to 19 per task), with execution timing and success rates differing substantially.
The resolve rate difference between \textsc{Prohibited} and \textsc{Unrestricted} is 1.25 percentage points on average, which is not statistically significant ($p > 0.05$)\revision{, and stays within $\pm 5$pp on every cell---including when the restriction is tool-enforced at the CLI level on Claude Code \textsc{Verified} ($-4$pp)}.
\textsc{Prohibited} mode consumes 56--62\% fewer tokens and 48--54\% less wall-clock time for Claude Code.
We find that execution has limited impact because (1) reproduction execution provides limited localization benefit, and (2) validation feedback contains significant noise (15--34\% environment errors).
These observations provide empirical insights into the role of execution in LLM-based repair.
Future work could further investigate the conditions under which execution feedback is most beneficial, develop agents that adapt execution strategies to task characteristics, and extend the analysis to other software engineering tasks.


\section*{Data Availability}
Our code and results are available at \url{https://github.com/mathieu0905/To_Run_Or_Not_To_Run}.



\bibliographystyle{ACM-Reference-Format}
\bibliography{main}

\end{document}